# Interface Effects on Tunneling Magnetoresistance in Organic Spintronics with Flexible Amine-Au Links

Narjes Gorjizadeh[1], Su Ying Quek[1,2,*]

[1] Institute of High Performance Computing, Agency for Science, Technology and Research, Singapore
[2] Department of Physics, National University of Singapore, S12-M01, 2 Science Drive 3, Singapore 117551
*To whom correspondence should be addressed: phyqsy@nus.edu.sg

**ABSTRACT**

Organic spintronics is a promising emerging field, but the sign of the tunneling magnetoresistance (TMR) is highly sensitive to interface effects, a crucial hindrance to applications. A key breakthrough in molecular electronics was the discovery of amine-Au link groups that give reproducible conductance. Using first principles calculations, we predict that amine-Au links give improved reproducibility in organic spintronics junctions with Au-covered Fe leads. The Au layers allow only states with *sp* character to tunnel into the molecule, and the flexibility of amine-Au links results in a narrow range of TMR for fixed number of Au layers. Even as the Au thickness changes, TMR remains positive as long as the number of Au layers is the same on both sides of the junction. Since the number of Au layers on Fe surfaces or Fe nanoparticles can now be experimentally controlled, amine-Au links provide a route towards robust TMR in organic spintronics.

**PACS numbers**
73.63.-b, 72.25.-b, 31.15.E-, 73.40.-c

## 1. Introduction

Organic molecules are promising alternatives to conventional semiconductors for spintronics applications. Besides low cost and mechanical flexibility, the weak spin-orbit interaction in organic systems can give rise to longer spin-coherence time and distances compared to inorganic semiconductors.[1,2] Chemical functionalization can also lead to magnetic centers within the molecules, with the potential for interesting spintronic effects.[3] Thanks to rapid progress in experimental techniques, spin injection into organic molecules is now a feasible task, typically leading to magnetoresistive responses of a few hundred percent at low temperature and a few percent at room temperature.[4-8] However, spin transfer between ferromagnetic contacts and organic molecules is a difficult procedure due to the large difference in their electronic structures and conductivities,[9] and the sensitivity to contact geometry[10, 11]. For example, the negative tunneling magnetoresistance (TMR; corresponding to lower resistances for the antiparallel configuration) in LSMO/Alq3/Co devices was once described as one of the most widely reproduced results in organic spintronics.[1] However, Fert et. al. later showed that these devices exhibited positive TMR for locally probed thin tunnel barriers, and proposed that the discrepancy arose from differences in interface hybridized states.[5] For spintronics applications, such as data storage, where the direction of stored spin is detected according to the sign of TMR, it is crucial that the sign of TMR be robust against any uncontrollable variations in junction geometry.

Since the interface is crucial to determining the resulting spin polarization and TMR in organic spintronics devices,[11, 12] systematic studies of interface-magnetoresistance relationships are of paramount importance. Single-molecule junctions, or self-assembled monolayers consisting of a single molecular layer, represent the smallest organic spintronic devices, and are a well-defined system for investigating interface-magnetoresistance effects. In recent years, experimentalists have succeeded in measuring the magnetoresistance of molecular-scale junctions,[13, 14] while the related field of molecular electronics has advanced considerably, with recent demonstrations of flexible molecular-scale electronic devices that remain stable over a thousand bending cycles.[15] Central to the progress in molecular electronics is the improved reproducibility of experiments on molecular-scale junctions,[16-18] a key step that paved the way for fundamental studies of charge transport across organic-inorganic interfaces. For organic spintronics to become truly viable, it is critical not only to understand interface-magnetoresistance effects, but also to find a system with reproducible magnetoresistance, and specifically, as an important first step, a robust sign of TMR.

The improved reproducibility of single-molecule charge transport measurements arose primarily from the discovery of chemical link groups, such as amine-Au link groups, that are selective but flexible.[16, 19-21] The amine group binds selectively to undercoordinated Au binding sites, and a large variation in molecular tilt angle and bonding configuration can be accommodated with a small range of Au-N-C bond angles and Au-N bond lengths.[19] The resulting conductance does not vary significantly with interface geometry; the measured conductance of benzene-diamine-Au junctions has been reproduced by different experimental groups,[17, 22] as well as by benchmark first principles calculations that include environment-dependent self-energy corrections in the energy alignments (DFT+$\Sigma$).[19, 23] Motivated by the success of amine-Au linkages in molecular electronics, we apply the DFT+$\Sigma$ approach to investigate whether amine-Au linkages can also be used to give reproducible magnetoresistance with Au-covered Fe electrodes. Specifically, we focus on exploring structure-magnetoresistance relationships in a junction with benzene-diamine (BDA; a prototypical organic molecule) bound *via* the amine groups to two Au-covered Fe electrodes.

In a recent review, Sanvito et al. proposed that one mechanism for spin filtering in molecular spintronics arises from the fact that the frontier molecular levels couple differently to different spin states, resulting in different electrode-induced renormalization and broadening effects.[10] This mechanism can be used to explain recent giant magnetoresistance effects in single phthalocyanine junctions.[14] Another recently proposed mechanism involves spin-dependent trapping of electrons at the organic/metal interface.[24] However, in recent experiments in which a non-magnetic tunneling barrier is added to improve reproducibility in organic spintronics,[6] the molecule is not bonded directly to the magnetic electrode and is thus not likely to be spin-polarized. Projected density of states on the molecule in the Fe-Au-BDA-Au-Fe junctions considered here also indicates that the BDA molecule is not spin-polarized. What, then, is the mechanism of spin transport in such systems?

## 2. Methods

The structures are optimized using density functional theory (DFT) with the PBE[25]

exchange-correlation functional as implemented in SIESTA.[26] All atoms in the molecule and Au layers, as well as in the top 4 Fe layers in the 6-layer Fe slabs are allowed to relax, with a force tolerance of 0.05 eV/Å. The Au lattice is rotated by 45° to match with Fe lattice. The computed lattice mismatch between Fe and Au is 2.6%. The lattice constant of relaxed Fe, 2.86Å, was used as the lattice constant of the Fe electrodes and the Fe-Au junctions. The distance between the electrodes is not constrained during geometry optimization (except for the stretched junction). In all cases, the amine group binds *via* the N lone pair to an undercoordinated atop Au site.

For the transport calculations, the system is divided into three parts: a scattering region and two magnetic leads. The scattering region in all the structures includes a molecule sandwiched between Au layers which are in contact with four atomic layers of Fe in contact with the periodic left and right Fe leads. The conductance is computed from the Landauer formula using the DFT+$\Sigma$[19, 23] approach as implemented in the scattering state transport code, SCARLET,[27] and the tunneling magetoresistance (TMR) effect is obtained as TMR=$(T_P-T_{AP})/\min(T_P,T_{AP})$, where $T_P$ and $T_{AP}$ are the total transmission in the parallel (P) and anti-parallel (AP) configurations of the leads, respectively. In the DFT+$\Sigma$ approach, inaccuracies in the DFT energy level alignment in the junction are corrected by adding to the scattering-state Hamiltonian a term of the form $\hat{\Sigma} = \sum_n \Sigma_n |\psi_n^{mol}\rangle\langle\psi_n^{mol}|$, where $|\psi_n^{mol}\rangle$ denotes the eigenstates of the molecular sub-Hamiltonian in the junction, and $\{\Sigma_n\}$ are the self-energy corrections for each molecular level (the DFT charge density is used as input in this 'one-shot' correction).[23] The self-energy correction is obtained using a physically-motivated parameter-free approach,[28] and consists of two parts: first, a 'bare' term, accounting for errors in the gas phase orbital energies, and second, an 'image-charge' term accounting for the effect of electrode polarization. Details of computing the self-energy follow reference 29. The density matrix is converged using a 2 × 2 $k_{//}$-mesh to sample the two-dimensional Brillouin Zone. However, a much denser $k_{//}$-mesh is required for convergence of the transmission – in this case, we use a 96 × 96 $k_{//}$-mesh, except for structure 7 for which we use a 128 × 128 $k_{//}$-mesh. This requirement for such a dense $k_{//}$-mesh is consistent with previous calculations on thiol-Ni junctions. The convergence of the $k_{//}$-mesh sampling is checked with a larger mesh of 128x128 (160x160 for structure 7) and has a deviation between 0-10% in transmission.

A double-ζ basis set is used for 6*s* orbitals while a single-ζ basis set is considered for 6*p* and 5*d* shells of Au during relaxation. Care is taken to choose Au basis sets that can reproduce the Au work function in a Au slab - an extra 7*s* orbital with single-ζ basis set is added to the surface Au atoms during transmission calculations.[30] The 4*s* and 3*d* orbitals of Fe are described by double-ζ basis sets while a single-ζ basis set is considered for 4*p* orbitals. Computed band structures of bulk Fe and Au are in very good agreement with those obtained by plane-wave method calculated by VASP.[31] These basis sets give the *s-d* hybridization gap of bulk Fe in good agreement with VASP, with a discrepancy of 1%. The interlayer distances between the surface Fe layers and Fe-Au layers are in good agreement with experiment[32] with errors of 1-14%. The work function of Fe (covered with Au ghost orbitals) is also in good agreement with VASP, with a discrepancy of 4%.

## 3. RESULTS AND DISCUSSION

For clarity, we begin our discussion with symmetric junctions that are repeated periodically in two dimensions perpendicular to the transport direction, typical models for single molecule junctions and self-assembled monolayers. Table I shows 8 different geometries of Fe-Au-BDA-Au-Fe junctions considered in this work. The bcc(100) Fe leads are covered by fcc(100) Au layers – this system has been realized experimentally due to its excellent lattice match, with its structure confirmed using ion scattering experiments.[33] The number of Au layers covering each lead ranges from 1 to 4.

The structures are optimized using density functional theory (DFT) with the PBE[25] exchange-correlation functional as implemented in SIESTA.[26] The conductance is computed from the Landauer formula using the DFT+$\Sigma$[19, 23] approach as implemented in the scattering state transport code, SCARLET.[27] The DFT+$\Sigma$ approach corrects for inaccuracies in the energy level alignment in the junction, taking into account experimentally observed image charge effects.[34]

### 3.1. Robust TMR

In all cases, we find that transmission occurs predominantly through the highest occupied molecular orbital, as is the case for Au-BDA-Au junctions.[19] The TMR values in Table 1 show that remarkably, all the structures considered here have the same sign of TMR. This is in contrast to thiol-Ni junctions, in which the TMR changes sign when the distance between electrodes changes by only 0.06 Å.[35] Furthermore, the predicted TMR is in general quite large compared to that computed for benzene-dithiol-Ni junctions.[36, 37] Comparing the transmission across all 8 structures, we see that in fact, the spin up transmission in the P configuration lies in a narrow range, with standard deviation of 0.6, which is one order of magnitude smaller than that of spin down transmission, i.e. 6.9. In the P configuration, the majority *sp* electrons from one Fe lead can couple directly to *sp* Au states, and tunnel through the BDA molecule to the other Au layer, and to the majority *sp* band of the other Fe lead. The narrow range of the spin up transmission in the P configuration thus arises from the fact that amine-Au link groups give rise to reproducible conductance in amine-Au junctions, where transmission takes place entirely *via sp* electrons. (There is negligible direct through-space tunneling in these geometries, unlike the case for BDA bound to flat Au(100) surfaces, a toy model discussed in the SI.)

For systems with 1 Au layer on each lead (structures 1-4), we consider different geometries (structures 1-4) involving different binding motifs, different molecular tilt angles and different distances between electrodes. Remarkably, not only is the sign of TMR robust, but also, the value of TMR falls within a reasonably narrow range of 47-77 % considering the variation of contact geometries involved. The spin up and spin down transmission for P and AP configurations are also very close in all the structures, except for structure 2 where the spin up transmission in P configuration and the transmission in AP configuration are larger, possibly because the Au-N-C angle is smaller (114° compared to 121-124° in other structures), resulting in a larger overlap between the N lone pair and the Au $s$ orbital.

### 3.2. Quantum Well States

For junctions with different number of Au layers, specifically, for structures 1, 5, 7 and 8, the TMR varies considerably (though still having the same sign). Similar oscillations, observed for the TMR in Fe-Au-MgO-Au-Fe junctions[38] and in giant magnetoresistance,[39] have been linked to quantum well states, which in this case arise from minority spin $sp$ electrons in Au being confined to the Au layer due to their inability to couple to the minority $d$ electrons in Fe. In particular, the TMR reported here appears to have a half-period oscillation, suggesting a full period of about 8 atomic layers, quite similar to that of quantum well states in Au in Au/Fe multilayers.[40, 41] We note that the large variation in TMR in these structures arises primarily from large changes in the spin down transmission in the P configuration, and moderate changes in the transmission in the AP configuration, an observation consistent with the role of minority spin quantum well states.

### 3.3. Interface States

Another interesting point to note is that for all structures except structure 8 in Table 1, we see that spin down transmission dominates in the P configuration. This is in contrast to the general rule that spin up $sp$ electrons conduct better than spin down $d$ electrons. In addition to minority spin quantum well states, which are less relevant for systems with a single Au adlayer,[41] another important factor that affects transmission is the existence of interfacial spin down states at the Fe layer closest to the Fe-Au interface. Since the predominantly *d* electrons for the minority spin in Fe cannot couple to the *sp* electrons of Au, spin down *d* electrons are accumulated at the interfacial Fe layer. This results in a larger density of spin down states at the interfacial Fe layer, as evident in the projected density of states (PDOS; Fig. 1). The relative dominance of spin up and spin down transmission will depend on a competition between availability of states and the ability of the states to couple across the junction.

Fig. 2 shows the transmission at $E_F$ as a function of $k_{//}$ for the P configuration. In general, the background transmission is larger for spin up (*e.g.* at the Gamma point, the spin up transmission is on average 1000 times larger than the spin down transmission). However, the maximum transmission for spin down is generally about one order of magnitude larger than that for spin up; significantly enhanced transmission is observed at specific $k_{//}$ (hot spots). By plotting the profiles of the conducting eigenchannels at hot spots and non-hot spots (Fig. 3), we see that the eigenchannel at a hot spot is characterized by peaks in the interfacial Fe layers at both sides of the junction, which couple across the molecule. Such hot spots have been predicted for other magnetic tunnel junctions,[42] including thiol-Ni junctions,[37] where they have been attributed to "near-resonant" transmission mediated by states at the interface between the magnetic electrodes and non-magnetic layers.[37, 42] It is helpful to note that while the interfacial states are clearly of predominant *d* character (Fig. 4c), *only* interface states that *also have sp* character can contribute to enhanced transmission (comparing Fig. 4a-c with Fig. 4d). Thus, hybridization between *sp* and *d* states in the interfacial Fe layer are important for coupling to Au. We expect that this physical picture is applicable also to other magnetic junctions where hot spots have been observed.[37, 42] Indeed, our modified Julliere model in 3.4 below quantitatively shows that the variation in interfacial

DOS is relatively small, and variations in TMR are almost entirely attributed to variations in the ability of spin up and spin down electrons to transmit across the junction.

The quantum well states discussed in 3.2 are particularly important in providing minority spin *sp* electrons in Au that allow the interfacial states to couple into the junction. The variation in spin down transmission for P configuration in structures 5-8 are related to the oscillations in TMR that arise from these quantum well states. In addition, we expect that the interface states also interact with the quantum well states in Au, resulting in relatively large changes in the spin down transmission for P configuration for structures 5-8. In structure 8, for example, the spin down transmission lacks hot spots and is very small, making transmission of this structure spin up dominant. This large variation of spin down transmission and change of spin dominancy due to the effect of quantum well states has also been observed in Fe-Au-MgO-Au-Fe junctions[38].

Since enhanced transmission occurs when interfacial Fe states on both sides of the junction couple across the molecule, the increased spin down transmission should be most significant for symmetric junctions in the P configuration, a conclusion that is in general consistent with our computed results (Table I). On the other hand, in the AP configuration, transmission is occurring from majority spin states on one side to minority spin states on the other, and vice versa, and thus, there are no hot spots (see SI Fig. 1). We also expect that the hot spots are less likely to exist in asymmetric junctions or junctions where periodic boundary conditions are broken.

To check the effect of symmetry, we calculate the TMR of an asymmetric structure with one Au layer but with an adatom contact on one side and a dimer contact on the other (Structure 9; Table II). We find that the TMR is reduced to ~12 % compared to the corresponding symmetric junctions due to the reduced matching of interface states at given $k_{//}$, resulting in a smaller spin down transmission in the P configuration. Importantly, however, the TMR remains positive. This is in contrast to an asymmetric thiol-Ni junction where asymmetric contacts change the sign of TMR.[43] However, it is important that to have the same number of Au layers on each side of the junction – for junctions with different Au thicknesses on either side of the junction, the TMR can become negative (see SI), an effect we attribute to complications due to quantum well states of different nature on both sides of the junction. Since the number of Au layers on Fe substrates and Fe nanoparticles can now be controlled[44], the negative TMR observed here can be avoided in practice. We further find that for rod-shaped leads instead of periodic leads, symmetric and asymmetric junctions with 1 Au layer on each lead have very similar TMR values, and in this case, the spin up transmission dominates (Structures 10-12; Table II). Thus, the effect of symmetry is likely to be less important when periodic boundary conditions are broken, and there are no hot spots.

**3.4. Modified Julliere Model**

We now discuss and summarize the above findings more quantitatively using a modified Julliere model. According to the original Julliere model, the conductance of P and AP configurations are proportional to the product of DOS of the two leads as follows: $G_P = G^{\uparrow\uparrow} + G^{\downarrow\downarrow}$ and $G_{AP} = G^{\uparrow\downarrow} + G^{\downarrow\uparrow}$, where $G^{\uparrow\uparrow} \propto \rho_1^\uparrow \rho_2^\uparrow$, $G^{\downarrow\downarrow} \propto \rho_1^\downarrow \rho_2^\downarrow$, $G^{\uparrow\downarrow} \propto \rho_1^\uparrow \rho_2^\downarrow$ and $G^{\downarrow\uparrow} \propto \rho_1^\downarrow \rho_2^\uparrow$. $G^{ij}$ is the conductance for spin-$i$-spin-$j$ configuration and $\rho_1^{i(j)}$ and $\rho_2^{i(j)}$ are the spin-polarized DOS of the two leads ($i,j=\uparrow,\downarrow$). Since the PDOS at the Fe/Au interface is quite different from the PDOS in the bulk Fe leads, it is appropriate to replace the lead DOS in the Julliere model with the PDOS of the Fe interfacial layer, similar to the procedure in Ref. 43.

In order to understand the effects of different Au layers, we first focus on structures 1, 5, 7 and 8. Defining $TMR^J=(G_P-G_{AP})/G_{AP}$, we obtain $TMR^J$ for structures 1, 5, 7, 8 as 159%, 231%, 204% and 152%, respectively. However, these values are quite different from the ab-initio results in Table I. This is because the model neglects important additional effects present in this study, such as the effect of quantum well states as well as different conductivities of spin up and spin down electrons. We thus modify the model by taking into account all the other important factors using a coefficient in front of the joint DOS: $T^{\uparrow\uparrow} = k^{\uparrow\uparrow}\rho_1^\uparrow \rho_2^\uparrow$, $T^{\downarrow\downarrow} = k^{\downarrow\downarrow}\rho_1^\downarrow \rho_2^\downarrow$, $T^{\uparrow\downarrow} = k^{\uparrow\downarrow}\rho_1^\uparrow \rho_2^\downarrow$ and $T^{\downarrow\uparrow} = k^{\downarrow\uparrow}\rho_1^\downarrow \rho_2^\uparrow$. Here, we use transmission (at $E_F$) instead of conductance because the two quantities are proportional according to the Landauer formula. First, we obtain best fit values for $k^{\uparrow\uparrow}$, $k^{\downarrow\downarrow}$, $k^{\uparrow\downarrow}$ and $k^{\downarrow\uparrow}$ to be $k^{\uparrow\uparrow}$=398.8, $k^{\downarrow\downarrow}$=30.0, $k^{\uparrow\downarrow}$=58.8 and $k^{\downarrow\uparrow}$=62.8. The one-order-of-magnitude larger value for $k^{\uparrow\uparrow}$ compared to $k^{\downarrow\downarrow}$ is due to larger conductivity of spin up *s* electrons than localized spin down *d* electrons. Using the modified Julliere model with the fitted parameters, the TMR$^J$ of structures 1, 5, 7 and 8 are 88%, 110%, 101% and 97%, respectively, in much better agreement with Table 1 than those obtained above, thus indicating that the additional factors capture the important physics in these systems. However, the corresponding variation in TMR$^J$ still has quite a different trend from the ab initio results.

To understand this, we compute $\frac{k^{\downarrow\downarrow}}{k^{\uparrow\uparrow}}$ separately for all structures. As shown in Fig. 5, the parameters $\frac{k^{\downarrow\downarrow}}{k^{\uparrow\uparrow}}$ do vary for different structures and the variation is in good agreement with variations in TMR. The variations are also compared with the ratio of spin down and spin up transmission in the P configuration ($\frac{T^{\downarrow\downarrow}}{T^{\uparrow\uparrow}}$) from table I. Since the TMR originates from differences in transmission of spin up and spin down electrons in the P configuration, the ratio $\frac{T^{\downarrow\downarrow}}{T^{\uparrow\uparrow}}$ has variations similar to TMR. According to our model, the ratio of spin down and spin up transmission is $\frac{T^{\downarrow\downarrow}}{T^{\uparrow\uparrow}} = \frac{k^{\downarrow\downarrow}}{k^{\uparrow\uparrow}} \frac{\rho^{\downarrow\downarrow}}{\rho^{\uparrow\uparrow}}$, where $\rho^{\downarrow\downarrow} = \rho_1^\downarrow \rho_2^\downarrow$ and $\rho^{\uparrow\uparrow} = \rho_1^\uparrow \rho_2^\uparrow$. Fig. 5 shows that $\frac{\rho^{\downarrow\downarrow}}{\rho^{\uparrow\uparrow}}$ in fact does not have significant variations, but the variation in $\frac{T^{\downarrow\downarrow}}{T^{\uparrow\uparrow}}$ comes from $\frac{k^{\downarrow\downarrow}}{k^{\uparrow\uparrow}}$, which indicates that variations in TMR result primarily from variations in the relative conductivity of spin up and spin down electrons and not from variations in interfacial PDOS.

Importantly, the variations in TMR and in $\frac{k^{\downarrow\downarrow}}{k^{\uparrow\uparrow}}$ are negligible for symmetric structures with the same number of Au layers but different molecule-Au contact geometries. Even for the asymmetric periodic structure (structure 9), the value of $\frac{k^{\downarrow\downarrow}}{k^{\uparrow\uparrow}}$ is reasonably close to those for the corresponding symmetric junctions (structures 1 and 3). Changing the Au thickness has a larger impact on $\frac{k^{\downarrow\downarrow}}{k^{\uparrow\uparrow}}$. For structures with rod-shaped leads, $\frac{k^{\downarrow\downarrow}}{k^{\uparrow\uparrow}}$ takes on a relatively narrow range, that is one order of magnitude smaller than those for periodic junctions, while $\frac{\rho^{\downarrow\downarrow}}{\rho^{\uparrow\uparrow}}$ is similar to that for periodic junctions. Interestingly, $\frac{k^{\downarrow\downarrow}}{k^{\uparrow\uparrow}} < 1$ for all structures considered here, even though spin down transmission dominates in structures 1-7. This indicates that the larger spin down transmission in these structures arises from the fact that $\frac{\rho^{\downarrow\downarrow}}{\rho^{\uparrow\uparrow}} > 10$, i.e. the interfacial spin down DOS is much larger than the interfacial spin up DOS. For structures 8-12, $\frac{\rho^{\downarrow\downarrow}}{\rho^{\uparrow\uparrow}}$ is still very large, but spin up transmission dominates because of the drop in $\frac{k^{\downarrow\downarrow}}{k^{\uparrow\uparrow}}$, i.e. because of the lack of hot spots in spin down transmission, which decreases $k^{\downarrow\downarrow}$.

## 4. CONCLUSIONS

We conclude by commenting on possible physical origins for why amine-Au links result in improved reproducibility in organic spintronics, compared to other link groups such as thiol-Ni links, where even the sign of TMR is not robust. In the case of molecules bonded directly to the magnetic leads, the magnetic proximity effect[35] for molecules next to the magnet would lead to spin polarization of the molecule, which in turn would be extremely sensitive to the metal-molecule interface.[10, 35, 43] By using a non-magnetic spacer Au layer between magnetic Fe leads and the molecule, we effectively allow only states with *sp*-character to tunnel across the molecule-metal interface. Amine-Au links are selective and flexible, and have been shown to give reproducible conductance in both theory and experiment, while in this work, we further show that TMR does not change significantly with changes in Au-molecule geometries. The role of hot spots in symmetric periodic junctions are interesting but not necessary for obtaining reasonably large positive TMR, and are not likely to be important in real experiments. In our proposed system, the major variation in TMR would arise from the effect of quantum well states in the Au layers, but this can be effectively controlled with present-day growth procedures. Thus, flexible amine-Au links provide a significant improvement over direct magnet-molecule links for reproducible organic spintronics.

**Acknowledgements**

We gratefully acknowledge H. J. Choi for providing access to the SCARLET code, as well as funding from A*STAR *via* the IHPC Independent Investigatorship, and computational resources from ACRC.

Table I. Transmission and TMR of different Fe-Au-BDA-Au-Fe junctions at the Fermi level.

| Structure index | Junction geometry BDA molecule: 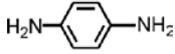 | | Fe leads | transmission (x10⁻³) | | TMR (%) |
|---|---|---|---|---|---|---|
| | | | | Spin UP | Spin DOWN | |
| 1 | 1 Au layer, Adatom contacts | 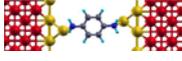 | P | 4.9 | 6.7 | 76 |
| | | | AP | 3.3 | 3.3 | |
| 2 | 1 Au layer, Adatom contacts (different angle) | 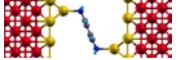 | P | 6.4 | 6.8 | 47 |
| | | | AP | 4.5 | 4.5 | |
| 3 | 1 Au layer, Dimer contacts | 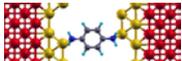 | P | 4.3 | 6.5 | 66 |
| | | | AP | 3.2 | 3.3 | |
| 4 | 1 Au layer, Adatom contacts; stretched by 0.2 Å | 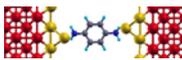 | P | 4.7 | 6.8 | 77 |
| | | | AP | 3.3 | 3.2 | |
| 5 | 2 Au layers, Adatom contacts | 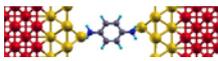 | P | 5.6 | 25 | 91 |
| | | | AP | 8.1 | 7.9 | |
| 6 | 2 Au layers, Dimer contacts | 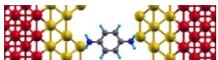 | P | 4.9 | 9.1 | 67 |
| | | | AP | 4.2 | 4.2 | |
| 7 | 3 Au layers, Adatom contacts | 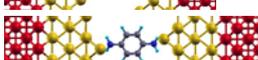 | P | 5.4 | 17 | 225 |
| | | | AP | 3.0 | 3.9 | |
| 8 | 4 Au layers, Adatom contacts | 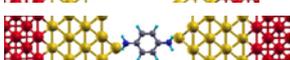 | P | 5.1 | 2.1 | 11 |
| | | | AP | 3.4 | 3.1 | |

Table II. Transmission and TMR of Fe-Au-BDA-Au-Fe junctions at the Fermi level for asymmetric periodic structure and the rod leads.

| Structure index | Junction geometry BDA molecule: 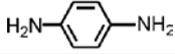 | | Fe leads | transmission (x10⁻³) | | TMR (%) |
|---|---|---|---|---|---|---|
| | | | | Spin UP | Spin DOWN | |
| 9 | Periodic, Asymmetric, 1 Au layer, Adatom-dimer contacts | 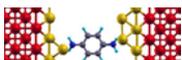 | P | 4.4 | 2.9 | 12 |
| | | | AP | 3.2 | 3.3 | |
| 10 | Rod, 1 Au layer, Adatom contacts | 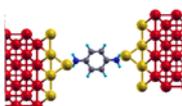 | P | 9.0 | 1.7 | 41 |
| | | | AP | 3.7 | 3.9 | |
| 11 | Rod, 1 Au layer, Dimer contacts | 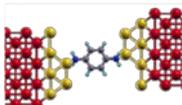 | P | 6.8 | 1.5 | 51 |
| | | | AP | 2.7 | 2.8 | |
| 12 | Rod, 1 Au layer, Asymmetric, Adatom-dimer contacts | 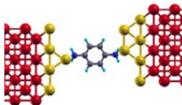 | P | 7.8 | 1.5 | 43 |
| | | | AP | 3.2 | 3.3 | |

Fig. 1 Projected Density of States (PDOS) of Fe layers for bulk Fe and for Fe layers near the Fe-Au interface for structure 1 (Fe layer 1 refers to the Fe layer closest to Au, and Fe layer 2 refers to the adjacent Fe layer). PDOS is shown for (a) spin up and (b) spin down carriers.

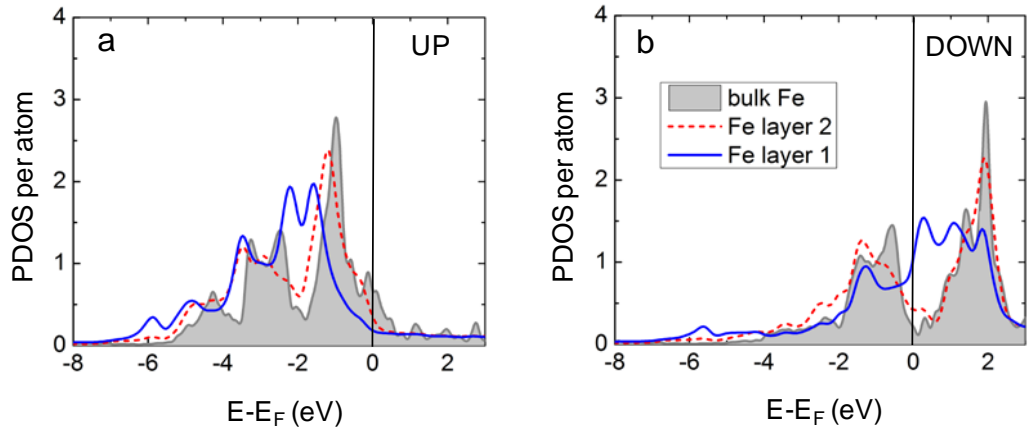

Fig. 2 Transmission at Fermi level versus $k_{//}$ for (a) spin up and (b) spin down carriers of structures of table I, for the P configuration. The structure index is shown in the middle of each pattern. The figures for structure 4 are very similar to those for structure 1 and are omitted from this plot.

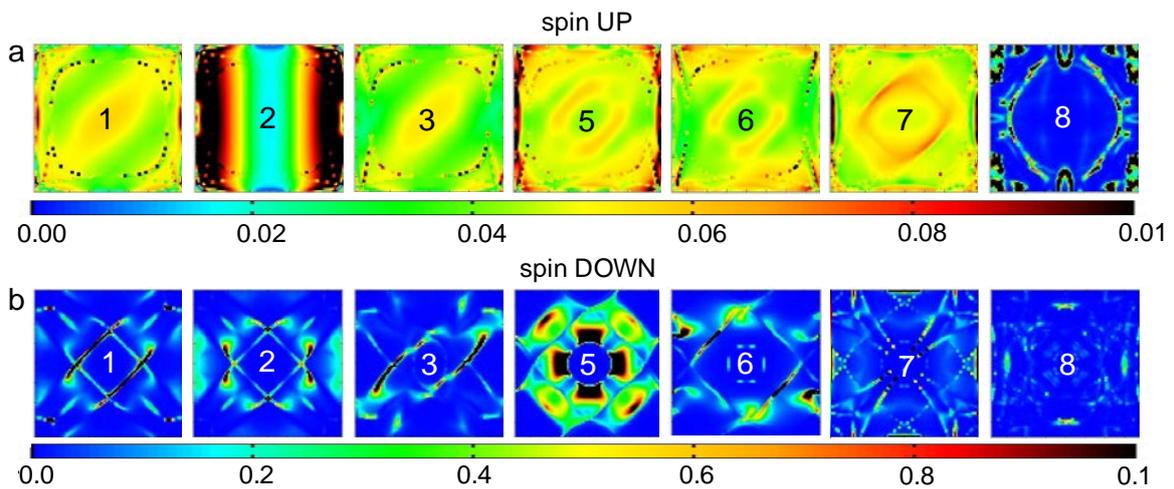

Fig. 3 Magnitude-squared of conducting eigenchannel wavefunction of structure 1, averaged over the *x-y* plane, plotted as a function of *z* (the transport direction; incident from small *z*), at (a) a hot spot for spin down, and (b) a non-hot spot. The inset shows the atomic structure with equivalent *z* coordinates to the plots. The positions of the interfacial Fe layer and the Au layers are defined by the solid and dashed vertical lines, respectively. The spin down wavefunction for the hot spot peaks at the interfacial Fe layers on both sides of the junction, indicating that the matching of interface states results in enhanced coupling and transmission.

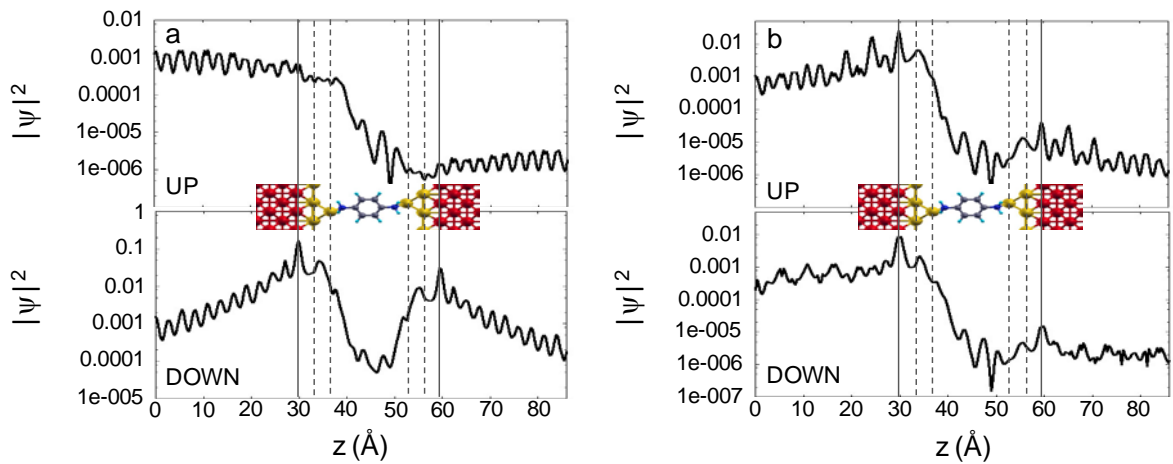

Fig. 4 (a-c) Spin down conducting eigenchannel wavefunction projected on interfacial Fe layer on the incident side, for (a) *s*, (b) *p* and (c) *d* orbitals versus $k_{//}$, in the P configuration of structure 1; (d) Transmission versus $k_{//}$ for spin down electrons of structure 1 in P configuration.

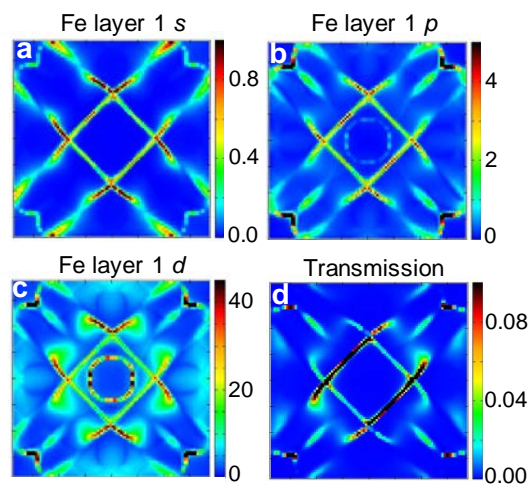

Fig. 5 TMR, ratio of spin down and spin up transmission in the P configuration ($\frac{T^{\downarrow\downarrow}}{T^{\uparrow\uparrow}}$), ratio of $\rho^{\downarrow\downarrow} = \rho_1^{\downarrow} \rho_2^{\downarrow}$ and $\rho^{\uparrow\uparrow} = \rho_1^{\uparrow} \rho_2^{\uparrow}$ ($\frac{\rho^{\downarrow\downarrow}}{\rho^{\uparrow\uparrow}}$), where $\rho_1^{i(j)}$ and $\rho_2^{i(j)}$ are the spin-polarized DOS of interfacial Fe layers at the two sides of the junction ($i,j=\uparrow,\downarrow$), and $\frac{k^{\downarrow\downarrow}}{k^{\uparrow\uparrow}}$ for symmetric junctions in table I. According to the modified Julliere model, transmission of spin up (spin down) electrons in the P configuration is proportional to $k^{\uparrow\uparrow}$ ($k^{\downarrow\downarrow}$) times DOS of the interfacial Fe layer at the two sides of the junction: $T^{\uparrow\uparrow} = k^{\uparrow\uparrow} \rho^{\uparrow\uparrow}$, $T^{\downarrow\downarrow} = k^{\downarrow\downarrow} \rho^{\downarrow\downarrow}$. Variation in TMR and $\frac{T^{\downarrow\downarrow}}{T^{\uparrow\uparrow}}$ results from the other effects that affect transmission ($\frac{k^{\downarrow\downarrow}}{k^{\uparrow\uparrow}}$) not DOS of interface states. 1L- 4L denote the number of Au layers (1 Au layer – 4 Au layers). The asymmetric structure 9 is denoted as 1L[a].

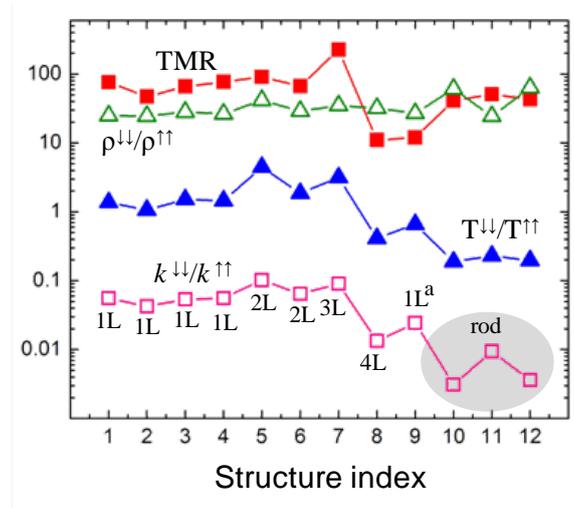